\documentclass[11pt]{article}
\usepackage{amsmath,amssymb,color,epsfig,cite,CJK}
\usepackage{graphicx}
\usepackage{subfigure}
\usepackage{setspace}
\usepackage{braket}

\usepackage{amsfonts}

\newcommand{\hoch}[1]{$\, ^{#1}$}

\makeatletter
\@addtoreset{equation}{section}
\makeatother

\newcommand{\be}{\begin{equation}}
\newcommand{\ee}{\end{equation}}
\newcommand{\bea}{\setlength\arraycolsep{2pt} \begin{eqnarray}}
\newcommand{\eea}{\end{eqnarray}}
\newcommand{\nn}{\nonumber}

\newcommand{\ra}{\rightarrow}

\newcommand{\bpm}{\begin{pmatrix}}
\newcommand{\epm}{\end{pmatrix}}

\def\ft#1#2{{\textstyle{\frac{\scriptstyle #1}{\scriptstyle #2} } }}
\def\fft#1#2{{\frac{#1}{#2}}}

\def\0{{\sst{(0)}}}
\def\1{{\sst{(1)}}}
\def\2{{\sst{(2)}}}
\def\3{{\sst{(3)}}}
\def\4{{\sst{(4)}}}
\def\5{{\sst{(5)}}}
\def\6{{\sst{(6)}}}
\def\7{{\sst{(7)}}}
\def\8{{\sst{(8)}}}
\def\sst#1{{\scriptscriptstyle #1}}
\def\oneone{\rlap 1\mkern4mu{\rm l}}
\def\ep{{\epsilon}}

\begin{document}


\begin{center}
{\large {\bf Quasi-topological Electromagnetism: Dark Energy, Dyonic Black Holes,\\
 Stable Photon Spheres and Hidden Electromagnetic Duality}}

\vspace{10pt}
Hai-Shan Liu\hoch{1,2}, Zhan-Feng Mai\hoch{1}, Yue-Zhou Li\hoch{1} and H. L\"u\hoch{1}

\vspace{10pt}

\hoch{1}{\it Center for Joint Quantum Studies and Department of Physics,\\
Tianjin University, Tianjin 300350, China}

\vspace{10pt}

\hoch{2}{\it Institute for Advanced Physics \& Mathematics,\\
Zhejiang University of Technology, Hangzhou 310023, China}

\vspace{40pt}

\underline{ABSTRACT}
\end{center}

We introduce the quasi-topological electromagnetism which is defined to be the squared norm of the topological 4-form $F\wedge F$. A salient property is that its energy-momentum tensor is of the isotropic perfect fluid with the pressure being precisely the opposite to its energy density.  It can thus provide a model for dark energy.  We study its application in both black hole physics and cosmology.  The quasi-topological term has no effect on the purely electric or magnetic Reissner-Nordstr\"om black holes, the dyonic solution is however completely modified.  We find that the dyonic black holes can have four real horizons. For suitable parameters, the black hole can admit as many as three photon spheres, with one being stable.  Another intriguing property is that although the quasi-topological term breaks the electromagnetic duality, the symmetry emerges in the on-shell action in the Wheeler-DeWitt patch.  In cosmology, we demonstrate that the quasi-topological term alone is equivalent to a cosmological constant, but the model provides a mechanism for the dark energy to couple with other types of matter. We present a concrete example of the quasi-topological electromagnetism coupled to a scalar field that admits the standard FLRW cosmological solutions.

\vfill {\footnotesize  hsliu.zju@gmail.com \ \ \ zhanfeng.mai@gmail.com\ \ \ liyuezhou@tju.edu.cn\ \ \  mrhonglu@gmail.com}

\thispagestyle{empty}

\pagebreak

\tableofcontents
\addtocontents{toc}{\protect\setcounter{tocdepth}{2}}

\newpage

\section{Introduction}

One important class of physical quantities observed in nature are described as form fields.  These include the Riemann curvature tensors, the Maxwell, or more generally Yang-Mills field strengths, all of which are 2-forms. The form fields have an intrigue feature that one can define the quantities that are independent of the metric or geometries of the spacetime, e.g.~
\be
\int {\rm tr} (R\wedge R)\,,\qquad \int F\wedge F\,,\qquad \cdots\,.
\ee
These terms give no contribution to the dynamics of a system, instead they describe its topology. However, dynamics can be built from these basic topological structures using the metric.  In fact, we may view the kinetic term of the Maxwell theory as the bilinear norm of $F$. Indeed $\int F$ measures the flux and is topological. This point of view leads to immediate higher-order generalizations to the standard electromagnetism.

The most famous nonlinear electromagnetism is the Born-Infeld theory \cite{Born:1934gh} whose Lagrangian can be written in a close form.  Perturbatively, it is an infinite series of invariant polynomials constructed from the Maxwell field strength, with the leading term being the Maxwell kinetic term.  In this paper, we consider nonlinear electromagnetism with polynomial invariants at some finite order, but constructed using the basic topological structures as the ingredients.  To be specific, for Maxwell field $F=dA$ in $D$ spacetime dimensions, we can construct a $(2k)$-form topological structure, as a wedge product of $F$ at the $k$'th order:
\be
V_{(2k)}=F\wedge F\wedge \cdots\wedge F\,,\qquad k\le [ D/2]\,.    \label{topexp}
\ee
Our quasi-topological polynomial invariants are the squared norms of $V_{(2k)}$'s,  namely
\be
U^{(k)} \sim |V_{(2k)}|^2\sim {*V_{(2k)}}\wedge V_{(2k)}\,.\label{squarenorm}
\ee
The simplest $k=1$ case gives rise to the usual kinetic term of the Maxwell theory.

The terminology ``quasi-topology'' was coined for two reasons.  One is for the nature of our construction.  The other is that for some special classes of Ans\"atze, such as any electrostatic system or light with global polarization, the quasi-topological polynomials give no contribution to the equations of motion.  This is analogous to quasi-topological curvature polynomial invariants in gravities, see e.g.~\cite{Oliva:2010eb,Myers:2010ru,Dehghani:2011vu,Li:2017ncu}.

One purpose of this paper is to study the effect of quasi-topological terms on the electromagnetism. We shall focus on four dimensions, where there is only one such term, corresponding to $k=2$. We couple the quasi-topological electromagnetism to Einstein gravity minimally and consider applications in both black hole physics and cosmology.

Although the quasi-topological term has no effect on equations of the purely electric or magnetic Reissner-Nordstr\"om (RN) black holes, they can nontrivially modify the dyonic solutions.  One unusual feature is that the dyonic solution can now have as many as four black hole horizons, instead of the usual two for the familiar RN black holes. For suitable parameters, we find that there can exist a stable photon sphere outside the dyonic black hole. Furthermore, we find the electromagnetic duality can emerge in the on-shell action, even though it is broken at the level of equations by the quasi-topological term.

One salient property of the quasi-topological term is that its energy-momentum tensor is of the perfect fluid, with isotropic pressure that is precisely the opposite to the energy density.  Thus the quasi-topological term can provide a candidate for dark energy.  The advantage of this dark energy model is that it is composite, built from a $U(1)$ field and hence it can provide interesting couplings to the dark or other forms of matter.
Therefore it may potentially realize the dark energy/dark matter interaction. (See a recent review \cite{Wang:2016lxa}.)

The paper is organized as follows.  In section \ref{sec:formalism}, we construct the quasi-topological terms of arbitrary order in general spacetime dimensions.  In section \ref{sec:d=4} and onwards, we focus on four dimensions.  We analyse the broken electromagnetic duality, energy conditions and dyonic particles. We demonstrate that the quasi-topological term can provide an on-shell spontaneous symmetry breaking mechanism.  In section \ref{sec:dbh}, we construct the exact solutions of general dyonic black holes carrying mass, electric and magnetic charges.  We analyse the global structure and demonstrate that as many as four black hole horizons can arise.  In section \ref{sec:ps}, we analyse the photon spheres created by the dyonic black holes and we find that a stable one can exist.  In section \ref{sec:hidden}, we demonstrate that although the electromagnetic duality is broken by the quasi-topological term, the duality can emerge in the on-shell action by adding some appropriate boundary terms.  In section \ref{sec:darke}, we study the application of the quasi-topological term in cosmology and show that it can be viewed as composite dark energy.  We conclude the paper and give further discussions in section \ref{sec:conclusion}. We give the dyonic black holes in general even dimensions in the appendix.

\section{Quasi-topological electromagnetism}
\label{sec:formalism}

\subsection{General construction}

In this paper, we consider higher-order extensions of the Maxwell theory.  We focus on the polynomial invariants of the field strength $F_{\mu\nu}=\partial_\mu A_\nu-\partial_\nu A_\mu$, but not of its derivatives.  The Lagrangian density of the usual Maxwell theory can be expressed as ${\cal L}=\sqrt{-g} L$, where $L$ is the quadratic invariant:
\be
L=-F^{\mu\nu} F_{\mu\nu} = -\delta_{\mu}^{[\rho} \delta_\nu^{\sigma]} F^{\mu\nu} F_{\rho\sigma}\,.
\ee
We consider a specific class of polynomial extensions where the Maxwell fields are contracted by the multi-index Kronecker delta, which is defined to have unit strength
\be
\delta^{\nu_1 \nu_2 \nu_3 \cdots \nu_{2k-1} \nu_{2k}}_{\mu_1 \mu_2 \cdots \mu_{2k-1} \mu_{2k}}=\delta^{[\nu_1}_{\mu_2} \delta^{\nu_2}_{\mu_1} \cdots \delta^{\nu_{2k-1}}_{\mu_{2k-1}} \delta^{\nu_{2k}]}_{\mu_{2k}}\,.
\ee
The polynomial invariant of the $(2k)$'th order is
\be
U^{(k)}=\frac{(2k)!}{(2k)!!}\delta^{\nu_1\nu_2\cdots\nu_{2k-1} \nu_{2k}}_{\mu_1\mu_2\cdots\mu_{2k-1} \mu_{2k}}
F^{\mu_1 \mu_2}\cdots F^{\mu_{2k-1}\mu_{2k}}F_{\nu_1 \nu_2 }\cdots F_{\nu_{2k-1}\nu_{2k}}\,.\label{exp-1}
\ee
It is thus clear that $U^{(k)}$ is the squared norm of the topological structure (\ref{topexp}).
Note that there is an alternative but equivalent expression, namely
\be
U^{(k)}= \fft{(2k)!}{(2k-1)!!}\delta^{\nu_1\nu_2\cdots\nu_{2k-1} \nu_{2k}}_{\mu_1\mu_2\cdots\mu_{2k-1} \mu_{2k}}
F^{\nu_1}_{\mu_1} \cdots F^{\nu_{2k}}_{\mu_{2k}}\,.\label{exp-2}
\ee
Note that in this paper we denote $F^\nu_\mu \equiv F^\nu{}_\mu = g^{\nu\rho} F_{\rho\mu}$.  However the expression (\ref{exp-1}) is more convenient to perform the variation principle for the Maxwell field. We define
\be
\widetilde F^{(k)\mu\nu} = \fft{\partial U^{(k)}}{\partial F_{\mu\nu}}=
\frac{2k\,(2k)!}{(2k)!!}\delta^{\mu\,\nu\,\nu_3\cdots\nu_{2k-1} \nu_{2k}}_{\mu_1\mu_2\mu_3\cdots\mu_{2k-1} \mu_{2k}}
F^{\mu_1 \mu_2}\cdots F^{\mu_{2k-1}\mu_{2k}}F_{\nu_3 \nu_4 }\cdots F_{\nu_{2k-1}\nu_{2k}}\,.
\ee
Then the $U^{(k)}$ term gives a contribution $\nabla_\mu \widetilde F^{(k)\mu\nu}$ to the Maxwell equation.  The totally antisymmetric property of the multi-index Kronecker delta implies that for given spacetime dimension $D$, we must have $k\le [D/2]$.  The contribution to the energy-momentum tensor from the $U^{(k)}$ term is given by
\bea
T^{(k)}_{\mu\nu} &=& \frac{2k (2k)!}{(2k)!!}\delta^{\nu_1\nu_2\cdots\nu_{2k-1} \nu_{2k}}_{\mu\mu_2\cdots\mu_{2k-1} \mu_{2k}}
F_{\nu}{}^{\mu_2} F^{\mu_3\mu_4}\cdots F^{\mu_{2k-1}\mu_{2k}}F_{\nu_1 \nu_2 }\cdots F_{\nu_{2k-1}\nu_{2k}}\nn\\
&&-\ft12 g_{\mu\nu} U^{(k)}\,.
\eea
Here $g_{\mu\nu}$ is the metric for a generic spacetime. Note that the right-hand side of the equation is automatically symmetric with the $\mu,\nu$ indices.

It is convenient to introduce the irreducible polynomial notations
\bea
F^{(2)} =F^{\mu}_\nu F^\nu_\mu = -F^2\,,\quad F^{(4)}=F^\mu_\nu F^\nu_\rho F^\rho_\sigma F^\sigma_\mu\,,\quad\cdots\,,\quad
F^{(2n)} = F^{\mu_1}_{\mu_2}F^{\mu_2}_{\mu_3} \cdots F^{\mu_{2n}}_{\mu_{1}}\,,
\eea
where $F^2\equiv F^{\mu\nu} F_{\mu\nu}$.  With these notations, we give some explicit low-lying examples:
\bea
U^{(1)} &=& -F^2\,,\nn\\
U^{(2)} &=& -2F^{(4)}+(F^2)^2\,,\nn\\
U^{(3)} &=& -8F^{(6)}-6F^2 F^{(4)}+(F^2)^3\,,\nn\\
U^{(4)} &=& -48F^{(8)}+32F^2F^{(6)}+12(F^{(4)})^2+12(F^2)^2F^{(4)}+(F^2)^4\,.
\eea
Our normalization for $U$ is that the last term has unit coefficient. Note that $F^{(2k+1)}$'s vanish identically owing to the fact that $F_{\mu\nu}=-F_{\nu\mu}$. Furthermore, if we define
\be
(F^{(k+1)})_{\mu\nu} \equiv F_{\mu\mu_1} F^{\mu_1}_ {\mu_2}\cdots F^{\mu_k}_\nu\,,
\ee
then we have
\be
(F^{(2k)})_{\mu\nu} = (F^{(2k)})_{\nu\mu}\,,\qquad (F^{(2k+1)})_{\mu\nu}=-(F^{(2k+1)})_{\nu\mu}\,.
\ee

\subsection{Quasi-topological properties}

Our construction in the previous subsection is analogous to that of the Euler densities in Riemannian geometry where the Riemann tensor polynomials are constructed by the contraction using the multi-index Kronecker delta. In fact, analogous construction involving both $F$ and the Riemann tensor leads to the Horndeski vector-tensor terms \cite{Horndeski:1976gi}. Unlike the Euler densities, our polynomials are not topological.  The reason for us to call these Maxwell polynomial invariants as quasi-topological is that for certain special class of ansatz, they give neither contribution to the Maxwell equation nor to the energy-momentum tensor. (This follows exactly the same definition of \cite{Li:2017ncu}.) As an example, we consider Minkowski spacetime in general $D$ dimensions, the metric in the Cartesian coordinates takes the form
\be
ds^2 = \eta_{\mu\nu} dx^\mu dx^\nu\,,
\ee
It is easy to see that for the ansatz of global polarization
\be
A_{\mu} =  \xi_\mu \phi(x)\,,\qquad \xi_\mu\, \hbox{is a constant vector,}\label{toposol}
\ee
the higher-order terms give no contribution to the Maxwell equation, and furthermore, their contributions to the energy-momentum tensor all vanish. In particular, this implies that the usual electromagnetic wave with constant global polarization (or a single photon) remains to be the solution in these higher-order extended theories.  The quasi-topological terms have no effect on the equation of any purely electric configuration either.

\section{Four dimensions}
\label{sec:d=4}

\subsection{Lagrangian, Hamiltonian and electromagnetic transformations}

In this paper, we focus on four dimensions, where there is only one quasi-topological term. Together with the Maxwell kinetic term, the theory is given by
\be
{\cal L}=\sqrt{-g}\Big(-\alpha_1 F^2 - \alpha_2 \big((F^2)^2 - 2F^{(4)}\big)\Big)\,,
\ee
where $\alpha_{1,2}$ are the two coupling constants. The case $\alpha_2=0$ gives the standard Maxwell theory, whilst $\alpha_1=0$ gives the quasi-topological electromagnetism. We shall present the formalism in a general curvature spacetime background, while many specific discussions will be given in the Minkowski spacetime.
The energy-momentum tensor of the system is given by
\bea
T_{\mu\nu} &=& \alpha_1 T^{(1)}_{\mu\nu} + \alpha_2 T^{(2)}_{\mu\nu}\,,\nn\\
T^{(1)}_{\mu\nu} &=& 2 F_{\mu\rho} F_{\nu}{}^\rho - \ft12 F^2 g_{\mu\nu}\,,\nn\\
T^{(2)}_{\mu\nu} &=& 4F^2 F_{\mu\rho} F_{\nu}{}^\rho - 8 F_{\mu\rho} F^{\rho}{}_\sigma
F^\sigma{}_\lambda F^\lambda{}_\nu - \ft12 \big((F^2)^2 - 2F^{(4)}\big) g_{\mu\nu}\,.\label{EMtensor}
\eea
It is no longer traceless, but
\be
T_\mu^\mu = 2\alpha_2 \Big((F^2)^2 - 2 F^{(4)}\Big)\,.
\ee
In general, the term $U^{(k)}$ gives rise to a traceless contribution to the energy-momentum tensor in $D=4k$ dimensions.

The Bianchi identity and the Maxwell equation of motion are
\bea
\hbox{BI}:&& \nabla_{[\mu} F_{\nu\rho]}=0\,,\qquad \hbox{EOM}:\ \  \nabla_\mu \widetilde F^{\mu\nu}=0\,,\nn\\ \widetilde F^{\mu\nu} &=& 4 \alpha_1 F^{\mu\nu}
+8\alpha_2 (F^2 F^{\mu\nu} - 2 F^{\mu\rho} F^{\sigma}{}_\rho F_{\sigma}{}^\nu)\,.\label{BIEOM}
\eea
The system of the equations are invariant under the interchanging of
\be
F_{\mu\nu}\qquad\leftrightarrow \qquad \ft12 \epsilon_{\mu\nu}{}^{\rho\sigma} \widetilde F_{\rho\sigma}\,.
\ee
However, the energy-momentum tensor $T_{\mu\nu}$ is not invariant under this electromagnetic transformation, except for $\alpha_2=0$.  On the other hand, we can perform an alternative version of the electromagnetic transformation, namely
\be
F_{\mu\nu}\qquad \leftrightarrow\qquad \ft12 \epsilon_{\mu\nu\rho\sigma} F^{\rho\sigma}\,,
\ee
The energy-momentum tensor is invariant, but not the set of equations.

To understand the issues, it is advantageous to write the field strength $F_{\mu\nu}$ in terms of electric and magnetic fields $(\vec E, \vec B)$, as
\be
F_{i0}=-F_{0i}=E_i\,,\qquad F_{ij} = \epsilon_{ijk} B_k\,.
\ee
The Lagrangian is now
\be
L = 2\alpha_1 (\vec E^2 - \vec B^2) + 8\alpha_2 (\vec E\cdot \vec B)^2\,.
\ee
The Hamiltonian $H=T^{00}$ is given by
\be
H=\alpha_1 (\vec E^2 + \vec B^2) + 4\alpha_2 (\vec E\cdot \vec B)^2\,.
\ee
Thus the situation with the ``electromagnetic duality'' becomes clear. Although the $\alpha_2$ term in both the Lagrangian and the Hamiltonian is invariant under $\vec E\leftrightarrow \vec B$, neither $\vec E$ and $\vec B$ is the fundamental field, and the transformation cannot be implemented locally on the fundamental field $A$.  Thus interchanging $\vec E$ and $\vec B$ is not consistent with the interchanging equations, namely the Bianchi identity and the equation of motion. From this point of view, the electromagnetic duality in the Maxwell theory is rather accidental.

We shall demonstrate in section \ref{sec:hidden} that there is a hidden electromagnetic duality in the on-shell action.

\subsection{Energy conditions}

From the discussion in the previous subsection, we see that the Hamiltonian is nonnegative provided that both $\alpha_1$ and $\alpha_2$ are nonnegative.
It is intriguing that the energy-momentum tensor $T^{(2)\mu\nu}$ is automatically diagonal like perfect fluid. Furthermore, the pressure $p$ is precisely the opposite of the energy density $\rho$, analogous to that of the cosmological constant or the dark energy:
\be
T^{(2)\mu\nu} =
\left(
  \begin{array}{cccc}
    \rho & 0 & 0 & 0 \\
    0 & p & 0 & 0 \\
    0 & 0 & p & 0 \\
    0 & 0 & 0 & p \\
  \end{array}
\right)\,,\qquad
\rho=-p=4(\vec E\cdot \vec B)^2\,.
\ee
Note that although the above discussion was based on the Minkowski spacetime, the conclusion holds in general curved spacetimes.

Since the Maxwell field with the standard kinetic term satisfies all the energy conditions \cite{Goldoni:2009zz}, it follows that for $\alpha_{1,2}>0$, the system satisfies the dominant energy condition, but not necessary the strong energy condition. Explicit examples will be given presently.

\subsection{A dyonic particle}

The $\alpha_2$-term has no dynamical effect on the solutions of the type (\ref{toposol}). In particular, it means that the point particles carrying either purely electric charge or purely magnetic charge give the same electrostatic or magnetic fields as in the Maxwell theory. The situation is different for the dyonic particles.  In the spatial spherical-polar coordinates, namely
\be
ds^2=-dt^2 + dr^2 + r^2 d\Omega_2^2\,,\qquad d\Omega_2^2 = d\theta^2 + \sin^2\theta d\varphi^2\,,
\ee
the ansatz of the Maxwell field for the dyonic particle located at the origin is
\be
A_\mu dx^\mu = \phi(r) dt + p \cos\theta d\varphi\,.
\ee
The equation of motion imply
\be
\phi'(r)=-\frac{q r^2}{\alpha_1 r^4+4\alpha_2 p^2}\,.\label{phiprime}
\ee
Thus we have
\be
{*(F\wedge F)}=\fft{2pq}{\alpha_1 r^4 + 4 \alpha_2 p^2}\,.
\ee
The electric potential can be easily integrated.  The answer depends on the sign of $\alpha_1\alpha_2$. As was discussed, unitarity of the theory requires that $\alpha_1\ge 0$ and $\alpha_2\ge 0$.  Thus we have three nontrivial cases
\bea
\alpha_1>0\,,\quad \alpha_2=0:&&\qquad \phi=\phi_0 + \fft{q}{\alpha_1 r}\,,\nn\\
\alpha_1=0\,,\quad \alpha_2>0:&&\qquad \phi=\phi_0 - \fft{q}{12\alpha_2 p^2} r^3\,,\nn\\
\alpha_1>0\,,\quad \alpha_2>0:&&\qquad \phi=\phi_0 + \fft{q}{\alpha_1 r} {}_2F_1[\ft14;1;\ft54, -\ft{4\alpha_2 p^2}{\alpha_1 r^4}]\,.
\eea
When $\alpha_2=0$, the electric potential is the standard one and it is divergent at the origin.  When $\alpha_1=0$, the electric potential is regular at the origin, but divergent asymptotically. When $\alpha_1\alpha_2\ne 0$, the electric potential is regular in the whole space, and its contribution to the energy-momentum tensor is thus regular. However, the existence of the magnetic monopole $p$ implies that the system is singular at the origin owing to the $\alpha_1$-term.  The on-shell Hamiltonian is
\be
H_{\rm on-shell}=\frac{q^2}{4 \alpha _2 p^2+\alpha _1 r^4}+\frac{\alpha _1 p^2}{r^4}\equiv\rho\,.
\ee
In other words, a magnetic monopole can regulate the divergence of the electric potential, but creates its own divergence. The energy momentum tensor in the diagonal vielbein basis is given by
\bea
&&T^{ab} = {\rm diag} \{\rho, p_r, p_2, p_3\}\,,\nn\\
p_r=-\rho\,,&&\qquad p_2=p_3=-\rho + 2\alpha_1 \Big(\frac{q^2 r^4}{\left(4 \alpha _2 p^2+\alpha _1 r^4\right){}^2}+\frac{p^2}{r^4}\Big)\,.
\eea
It can be easily demonstrated that the dyonic particle satisfies the null, weak and dominant energy condition,
but not necessary the strong condition.

The energy-momentum tensor is in general not constant; however, when $\alpha_1=0$, it is.  We find
\be
T^{ab}=\Lambda_{\rm eff}\, {\rm diag} \{1,-1,-1,-1\} \,,\qquad \Lambda_{\rm eff}=\fft{q^2}{4\alpha_2 p^2}>0\,.
\label{Lambdaeff}
\ee
Thus for the $\alpha_1=0$ case, the energy-momentum tensor has the same effect of a positive cosmological constant $\Lambda_{\rm eff}$. We shall come back to this point in section \ref{sec:darke}.

Finally we note that when $\alpha_1\alpha_2<0$, there is a singularity at $r_*$, given by
\be
r_*=\Big(-\fft{4\alpha_2 p^2}{\alpha_1}\Big)^{\fft14}\,.
\ee
The solution now becomes ($\alpha_1\alpha_2<0$):
\bea
r>r_*: &&\qquad \phi=\phi_0+
    \fft{q}{\alpha_1 r} {}_2F_1[\ft14;1;\ft54, \ft{r_*^4}{ r^4}]\,,\nn\\
0\le r< r^*: &&\qquad \phi=\phi_0
  -\ft{q r^3}{3\alpha_1 r_*^4} {}_2F_1[\ft34;1;\ft74, \ft{r^4}{r_*^4}]\,.
\eea
We shall not discuss this case further in this paper.

\subsection{An on-shell spontaneous symmetry breaking mechanism}

We consider a simple model of a massless scalar field $\phi$ coupled to the quasi-topological electromagnetism, with no Maxwell kinetic term. The Lagrangian in the Minkowski spacetime is
\be
{\cal L}= -\ft12 (\partial\phi)^2 -\alpha_2 \big((F^2)^2 - 2F^{(4)}\big) - \ft12 \beta \phi^2 \varepsilon^{\mu\nu\rho\sigma}F_{\mu\nu} F_{\rho\sigma}\,.\label{phiF}
\ee
where $\varepsilon^{\mu\nu\rho\sigma}$ is the totally antisymmetric tensor with $\varepsilon^{0123}=1$. Note that the sign choice of last term is inessential, and hence we choose without loss of generality that $\beta$ is positive. Defining ${*(F\wedge F)}=2\psi$, the equation of motion for $A$ implies that
\be
\psi = \lambda \big(1 + \fft{\beta}{4\alpha_2\lambda}\phi^2\big)\,,
\ee
where $\lambda$ is an integration constant, measuring the constant flux of $F\wedge F$.  The scalar equation is now given by
\be
\Box \phi= \fft{\partial V_{\rm eff}}{\partial\phi}\,,\qquad V_{\rm eff}=8\alpha_2\lambda^2 \big(1 + \fft{\beta}{4\alpha_2\lambda}\phi^2\big)^2\,.
\ee
Thus the effective Lagrangian for (\ref{phiF}) is simply a scalar theory with ${\cal L}_{\rm eff}=-\fft12 (\partial\phi)^2 - V_{\rm eff}$.  Thus although we start with a free scalar, turning on the constant flux $\lambda$ generates an effective mass, with $\mu^2=8\beta \lambda$, for the scalar.

      For positive flux $\lambda$, $\phi=0$ is the vacuum and we have an effective positive cosmological constant $\Lambda_{\rm eff}=4\alpha_2\lambda^2$. On the other hand, when the flux $\lambda<0$, $\phi=0$ is locally maximum, and the true vacuum is located at $\phi=\pm 2\sqrt{\alpha_2 (-\lambda)/\beta}$, where the effective cosmological constant vanishes.  Thus we see that the quasi-topological mechanism can provide an on-shell parameter (or integration constant) $\lambda$ for the spontaneous symmetry breaking. We demonstrated this using a real scalar; the generalization to the complex scalar is straightforward.

\section{Dyonic black holes}
\label{sec:dbh}

\subsection{Minimally-coupled to gravity}

We now consider gravity where the electromagnetism is minimally coupled. We also introduce a bare cosmological constant $\Lambda_0$ for generality. The Lagrangian is
\be
{\cal L}=\sqrt{-g}\Big(R-2\Lambda_0-\alpha_1 F^2-\alpha_2((F^2)^2-2F^{(4)})\Big)\,.\label{d4lag}
\ee
The bulk action is given by $I=1/(16\pi) \int d^4 x {\cal L}$. The Bianchi identity and the Maxwell equation take the same forms as in (\ref{BIEOM}), but now in general curved spacetime.  The Einstein field equation is
\be
R_{\mu\nu} - \ft12 R g_{\mu\nu} + \Lambda_0 g_{\mu\nu} = T_{\mu\nu}\,,
\ee
where $T_{\mu\nu}$ is given by (\ref{EMtensor}).  We also assume that both coupling constants $\alpha_1$ and $\alpha_2$ are nonnegative.

It is worth mentioning that the $\alpha_2$-term was discovered when we were looking for the quartic polynomials that would have no effect on the equations for the RN black hole carrying electric charges only.
The $\alpha_2$-term turns out to be the unique solution.

\subsection{${\cal M}_2\times S^2$ vacua}

Here we shall focus on having vanishing bare cosmological constant, i.e. $\Lambda_0=0$.  We are interested in vacuum solutions of the type
\be
ds^2 = ds_2^2 + d\Sigma_2^2\,,\qquad F=q \epsilon_2 + p \Sigma_2\,,
\ee
where $ds_2^2$ and $d\Sigma_2^2$ are two dimensional Einstein spaces with Lorentzian and Euclidean signatures respectively and $\epsilon_2$ and $\Sigma_2$ are the the respective volume 2-forms.  Assuming that $R_{\mu\nu}=\lambda_1 g_{\mu\nu}$ for $ds_2^2$ and $R_{ij}=\lambda_2 g_{ij}$ for $d\Sigma_2^2$, we find
\be
\lambda_1=-\alpha_1 (q^2 + p^2) + 4\alpha_2 q^2 p^2\,,\qquad
\lambda_2=\alpha_1 (q^2 + p^2) + 4\alpha_2 q^2 p^2\,.
\ee
Thus we see that the Euclidean signature space $d\Sigma_2^2$ must be a 2-sphere, while the Lorentzian signatured $ds_2^2$ can be AdS$_2$, or (Mink)$_2$ or dS$_2$, corresponding $\lambda_1<0, =0$ or $>0$ respectively.

\subsection{Dyonic black holes and the horizon structures}
\label{subsec:dbh}

\subsubsection{The solutions}

We find that the theory admits exact solutions of general spherically-symmetric and static dyonic black holes:
\bea
ds^2 = - f\, dt^2 + \fft{dr^2}{f} + r^2 d\Omega_{2,\epsilon}^2\,,\qquad
F=-\phi'(r) dt\wedge dr + p\, \Omega_{2,\epsilon}\,,
\eea
where $\epsilon=1,0,-1$, for which the metric $d\Omega_{2,\epsilon}^2$ describes a sphere, torus and hyperbolic 2-space respectively.  The electric potential $\phi$ satisfies (\ref{phiprime}) and the blackening factor $f$ satisfies
\be
f'+\frac{f-\epsilon}{r}+\frac{\alpha_1 p^2}{r^3}+\frac{q^2 r}{\alpha_1 r^4+4\alpha_2 p^2}+\Lambda_0 r=0\,.
\ee
The solution can be expressed as
\be
f(r)=-\ft13\Lambda_0 r^2+\ep -\frac{2M}{r}+\frac{\alpha_1 p^2}{r^2}+\frac{q^2}{\alpha_1 r^2} \, _2F_1\Big(\frac{1}{4},1;\frac{5}{4};-\frac{4 p^2 \alpha_2}{r^4 \alpha_1}\Big)\,.\label{fsol}
\ee
The general solution carries three integration constants, $(M,q,p)$, associated with the mass, electric and magnetic charges. As we have discussed in section \ref{sec:d=4}, the system satisfies at least the dominant energy condition when both $\alpha_1$ and $\alpha_2$ are positive. For the black hole solutions, we find
\bea
\rho &=&\Lambda_0 + \frac{q^2}{4 \alpha _2 p^2+\alpha _1 r^4}+\frac{\alpha _1 p^2}{r^4}\,,\qquad
\rho+p_1 =0\,,\nn\\
\rho+p_{2,3} &=& \frac{2 \alpha _1 q^2 r^4}{\left(4 \alpha _2 p^2+\alpha _1 r^4\right){}^2}+\frac{2 \alpha _1 p^2}{r^4}\,,\nn\\
\rho-p_{2,3} &=& 2\Lambda_0 + \fft{8\alpha_2 q^2p^2}{(\alpha_1 r^4 + 4\alpha_2 p^2)^2}\,,
\eea
where $p_1, p_{2,3}$ are the pressure in the the radial and 2-sphere directions respectively.  Thus we see that the black holes indeed satisfies the dominant energy condition if $\Lambda_0\ge 0$, including the asymptotically flat solutions.  The strong energy condition however can be violated since
\be
\rho + p_1 + p_2 + p_3 = -2\Lambda_0 + \fft{2\alpha_1 p^2}{r^4} - \fft{2q^2(4\alpha_2 p^2 - \alpha_1 r^4)}{(4\alpha_2 p^2 + \alpha_1 r^4)^2}\,.
\ee
This is not surprising since the quasi-topological term, which behaves like dark energy, will violate the strong energy condition.  Of course, the above only demonstrates that the strong energy condition can be violated by some black holes, but not by all.

\subsubsection{Special limits}

The hypergeometric function in (\ref{fsol}) becomes degenerated to polynomials in the following special cases:
\bea
p=0:&&\qquad f= -\ft13 \Lambda_0 r^2 + \epsilon - \fft{2M}{r} + \fft{q^2}{\alpha_1 r^2}\,,\nn\\
q=0:&&\qquad f=-\ft13 \Lambda_0 r^2 + \epsilon - \fft{2M}{r} + \fft{\alpha_1 p^2}{r^2}\,,\nn\\
\alpha_2=0:&&\qquad f=-\ft13\Lambda_0 r^2 + \epsilon - \fft{2M}{r^2} +
\fft{q^2}{\alpha_1 r^2} + \fft{\alpha_1 p^2}{r^2}\,,\nn\\
\alpha_1=0:&&\qquad f=-\ft13(\Lambda_0 + \Lambda_{\rm eff})\, r^2 + \epsilon - \fft{2M}{r}\,.\label{4spcases}
\eea
where $\Lambda_{\rm eff}$ is given by (\ref{Lambdaeff}). The first two cases are the reason we call the $\alpha_2$-term quasi-topological. Note that in taking the $\alpha_1=0$ limit, the divergent term can be absorbed by redefining the mass parameter $M$.  The geometries of all these special solutions are those of RN black holes or simply the Schwarzschild one, and their global structures have been well studied.

\subsubsection{Global analysis}

We thus focus on the case with $\alpha_1\alpha_2\ne 0$.  Asymptotically at large $r$, the blackening factor behaves as
\be
f=-\ft13\Lambda_0 r^2 + \epsilon - \fft{2M}{r} +\fft{q^2}{\alpha_1 r^2} + \fft{\alpha_1 p^2}{r^2}
-\frac{4 \alpha _2 p^2 q^2}{5 \alpha _1^2 r^6} + \frac{16 \alpha _2^2 p^4 q^2}{9 \alpha _1^3 r^{10}}+\cdots\,.
\ee
On the other hand, at the vicinity of the black hole curvature singularity $r=0$, we have
\be
f=\frac{\alpha _1 p^2}{r^2}+\frac{\frac{\pi  q^2}{4 \alpha _1^{3/4} \sqrt[4]{\alpha _2} \sqrt{p}}-2 M}{r}+1 -\left(\frac{\Lambda_0}{3}+\frac{q^2}{12 \alpha _2 p^2}\right)r^2 + \cdots\,.
\ee
Assuming that $\Lambda_0=0$ and $\epsilon=1$ so that the metric is asymptotically flat, then if there are horizons, there must be an even number of them.  This is analogous to the RN black hole, which has in general two horizons. We find that for suitable parameters, there can be four horizons. Note that multi-horizon solutions were also constructed in \cite{Gao:2017vqv} in the reverse-engineered $f(F^2)$ theories. These theories typically violate the null-energy condition and furthermore the locations of the horizons are the parameters of the theory rather than the integration constants of the solutions.

To understand the multiple horizon structures, it is useful to define a function $W= r f+2M$, which must be positive for $r>0$ and satisfies
\be
W'=1 -\frac{\alpha _1 p^2}{r^2}-\frac{q^2}{\alpha _1 r^2 \left(1+\frac{4 \alpha _2 p^2}{\alpha _1 r^4}\right)}\,.\label{wprime}
\ee
Note that we are considering the case where the constants $(\alpha_1,\alpha_2,q,p)$ are all positive.
Thus we see that $W'=0$ is a cubic equation of $r^2$, which has either one real root or three real roots for $r^2$.  It is clear that any real root for $W'=0$ must be positive for $r^2$, it implies that $W'=0$ has either one positive root or three positive roots of the coordinate $r$.  When there is only one positive root for $r$, then the function $W$ has one minimum $2M_0>0$.  It follows that for $M> M_{\rm 0}$ there are two horizons and they coalesce when $M=M_0$. The curvature singularity becomes naked when for $M<M_0$.

When $W$ has three positive roots, it has three positive local extrema $(2M_1,2M_2,2M_3)$, as $r$ runs from zero to infinity, where $(2M_1,2M_3)$ are two minima and $2M_2$ is the maximum.  Define $M_+={\rm max}(M_1,M_3)$ and $M_-={\rm min}(M_1,M_3)$, we have $0<M_-\le M_+ <M_3$.  The horizon structures are then dictated by the following rules:
\bea
M< M_-:&&\quad\hbox{no horizon with naked singularity};\nn\\
M_-\le M\le M_+ :&&\quad\hbox{two horizons};\nn\\
M_+< M < M_2 :&&\quad \hbox{four horizons};\nn\\
M_2<M:&&\quad \hbox{two horizons}.\label{horizonrule}
\eea
Note that in the above when an equality is saturated, two horizons coalesce. If $M=M_-$ or $M=M_+$,
the near-horizon geometry is AdS$_2\times S^2$; if $M=M_2$, there is a dS$_2\times S^2$ in the interior.  If $M=M_+=M_-$, then the black hole has two extremal horizons.

We now illustrate the above discussion with explicit examples.  Since we only consider the cases with $\alpha_1>0$, we can set $\alpha_1=1$ without loss of generality.  Furthermore for simplicity, we shall treat $\alpha_2$ as a variable rather than a fixed coupling constant; otherwise, it can be very tedious to solve
algebraic equation involving the hypergeometric functions.  Assuming that $W'$ in (\ref{wprime}) has three roots $(x_1,x_2,x_3)$ for $r^2$, with $0<x_1<x_2<x_3$.  We can express $(q,p,\alpha_2)$ in terms of these roots:
\bea
q^2&=&\frac{\left(x_1+x_2\right) \left(x_1+x_3\right) \left(x_2+x_3\right)}{x_1 x_2+x_3 x_2+x_1 x_3}\,,\qquad
p^2=\frac{x_1 x_2 x_3}{x_1 x_2+x_3 x_2+x_1 x_3}\,,\nn\\
\alpha_2&=&\frac{\left(x_1 x_2+x_3 x_2+x_1 x_3\right){}^2}{4 x_1 x_2 x_3}\,.
\eea
The thresholds $M_{0,1,2}$ can be read off from $W$ by substituting these roots.

\bigskip
\noindent{\bf Case 1: black holes with at most two horizons}
\bigskip

In this case, only one of the three $x_i$'s is real.  For a concrete example, we consider
\be
(x_1,x_2,x_3)=(1+{\rm i}, (1-{\rm i}), 1)\,,\qquad
\rightarrow\qquad (q^2,p^2,\alpha_2)\rightarrow (\ft52,\ft12,2)\,.
\ee
Thus we have
\be
M_0=\ft{3}{4} + \ft{5}{4} \, _2F_1[\ft{1}{4},1;\ft{5}{4};-4]\sim 1.63724\,.
\ee
The solution describes a black hole with two horizons when $M> M_0$, and it becomes extremal when $M=M_0$. The solution suffers from having naked curvature singularity at $r=0$ when $M<M_0$.  Explicitly, we have
\be
f=1-\frac{2 M}{r}+\frac{1}{2 r^2} \big(1+5\, _2F_1[\ft{1}{4},1;\ft{5}{4};-\ft{4}{r^4}]\big)\,,\qquad M\ge M_0\,.\label{case1bh}
\ee
Note that in this category, the function $f$ is monotonically increasing outside the outer horizon. This black hole, as in the case of RN black holes, satisfies the strong energy condition as well as the dominant energy condition.

\bigskip
\noindent{\bf Case 2a: black holes with at most four horizons}
\bigskip

We now consider the cases where all the roots are real and hence positive. Assuming that $x_1<x_2<x_3$, we first focus on the case with $M_3< M_1$.  An explicit example is provided with
\be
(x_1,x_2,x_3)=(1,7,36)\,,\qquad \rightarrow\qquad (q^2,p^2,\alpha_2)=(\ft{12728}{295},\ft{252}{295},\ft{87025}{1008})\,,
\ee
for which, we have $(M_1,M_2,M_3)=(6.6845 , 6.8437,6.5209)$. The solution is given by
\be
f=1 - \fft{2M}{r} + \fft{4}{295r^2} \big( 63 + 3182\, _2F_1[\ft{1}{4},1;\ft{5}{4};-\ft{295}{r^4}]\big)\,.
\label{case2abh}
\ee
The counting of the numbers of possible horizons follows the rule (\ref{horizonrule}).  In particular, four real horizons can emerge.  For example, when $M=6.7$, the four horizons are located at
\be
r_1=7.9283\,,\qquad r_2=4.0670\,,\qquad r_3=1.2413\,,\qquad r_4=0.8156\,.\label{case2a4h}
\ee
Note that in this category, the function $f$ is also monotonically increasing outside the outer horizon.
This solution satisfies only the dominant energy condition, but not the strong energy condition.

\bigskip
\noindent{\bf Case 2b: black holes with at most four horizons}
\bigskip

We now consider the case with $M_3> M_1$.  A concrete example is
\be
(x_1,x_2,x_3)=(1,11,36)\,,\qquad \rightarrow\qquad (q^2,p^2,\alpha_2)=
(\ft{20868}{443},\ft{396}{443},\ft{196249}{1584})\,,
\ee
with $(M_1,M_2,M_3)=(6.7730, 6.9135, 6.6316)$. The solution is given by
\be
f=1-\fft{2M}{r} + \fft{12}{443r^2} \big(33 + 1739 \, _2F_1[\ft{1}{4},1;\ft{5}{4};-\ft{443}{r^4}]\big)\,.
\label{case2bbh}
\ee
This solution satisfies the dominant energy condition, but not the strong energy condition.
What distinguishes this case with the earlier case 2a is that when $M_1\le M<M_3$, the black hole has two horizons, but the function $f$ is not monotonically increasing outside the outer horizon.  For example, we choose $M=6.7$, and in this case, the outer and inner horizons are located at
\be
r_1=1.5498\,,\qquad r_2=0.6682\,.\label{case2b2h}
\ee
Interestingly, in the outer region $r>r_1$, the function $f$ has a wiggle with one local maximum and one local minimum, located at $r=2.848$ with $f=0.138826$ and $r=6.1421$ with $f=0.0241$ respectively.

What is striking about this case is that the Newtonian potential for this black hole has a stable equilibrium at $r=6.1421$, the distance about four times the outer horizon radius.  In the region of $r\in (2.848,6.1421)$, the gravity force is repulsive instead of being attractive. The previously known examples of the black hole repulsion all involve the repulsive scalar charges, see e.g.~\cite{Horne:1992bi,Lee:1994sk,
Gibbons:1994ff,Lee:1991qs,Zhao:2018wkl,Lu:2019icm}. The situation here is very different since there is no scalar in the theory.

The existence of a static equilibrium for a massive particle owing to the extra wiggle of $f$ is suggestive of the possibility of a stable photon sphere, which we shall discuss in the next section.

\bigskip
\noindent{\bf Case 3: black holes with two extremal horizons}
\bigskip

It is also of interest to present a dyonic black hole with two extremal horizons.  Let us take the parameters to be
\bea
&&\alpha_2=\fft{81}{4p^2 (9-10 p^2)}\,,\qquad
q^2 = \fft{10 (9- p^2) (1 -  p^2)}{9-10 p^2}\,,\nn\\
&&M=\ft12 (1 +  p^2) + \fft{q^2}{2} {}_2F_1[\ft14,1;\ft54;-\ft{81}{9-10p^2}]\,.
\label{case3bh}
\eea
We find that the black hole has two extremal horizons at $r_1=3=r_2$ and $r_3=1=r_4$, provided that $p^2 =0.71289\cdots$. In this case, we have mass $M=3.56111$. Again this black hole satisfies the dominant energy condition, but not strong energy condition.

Thus for the asymptotically flat solutions $(\Lambda_0=0)$, we see that in the case where $(q,p,\alpha_2)$ are given such that adjusting $M$ can give at most two horizons, the back holes satisfy the strong energy condition as well as the dominant energy condition.  When adjusting $M$ can give as many as four horizons, the black holes satisfy the dominant, but not the strong energy condition. The existence of four black hole horizons implies that an object falling into the outer horizon can enter a new ``habitable'' world sandwiched between a black hole horizon and a cosmic horizon.

\subsection{Black hole thermodynamics}

From the asymptotic behavior of the blackening factor $f$, it is straightforward to see that the dyonic black hole has mass $M$.  The electric and magnetic charges can be read off from the equations of motion and the Bianchi identity, namely
\be
Q_e = \fft{1}{4\pi} \int \widetilde F^{0r} = q\,,\qquad
Q_m = \fft{1}{4\alpha_1 \pi} \int F = \fft{p}{\alpha_1}\,.
\ee
We see clearly that the electric and magnetic charges enter the metric asymmetrically and hence the electromagnetic duality breaks down by the $\alpha_2$ term.

Assuming that the outer horizon is located at $r_+$, we find that the Hawking temperature and the entropy are
\be
T=-\frac{q^2 r_+}{4 \pi  \left(4 \alpha _2 p^2+\alpha _1 r_+^4\right)}-\frac{\alpha _1 p^2}{4 \pi  r_+^3}-\frac{\Lambda _0 r_+}{4 \pi }+\frac{\epsilon}{4 \pi  r_+}\,,\qquad S=\pi r_+^2\,.
\ee
The electric and magnetic potentials are
\bea
\Phi_e&=&\int_{r_+}^\infty a'=\frac{q \, _2F_1\left(\frac{1}{4},1;\frac{5}{4};-\frac{4 p^2 \alpha _2}{r_+^4 \alpha _1}\right)}{\alpha _1 r_+}\,,\nn\\
\Phi_m &=& -\frac{q^2 \, _2F_1\left(\frac{1}{4},1;\frac{5}{4};-\frac{4 p^2 \alpha _2}{r_+^4 \alpha _1}\right)}{4 p r_+}+\frac{\alpha _1 q^2 r_+^3}{4 p \left(4 \alpha _2 p^2+\alpha _1 r_+^4\right)}+\frac{\alpha _1^2 p}{r_+}\,.
\eea
It is now straightforward to verify that the first law of thermodynamics
\be
dM=TdS + \Phi_e dQ + \Phi_m dQ_m + \ft43 \pi r_+^3 d(-\ft{\Lambda_0}{8\pi})\,.\label{firstlaw}
\ee
Note that the last term is treating the a negative cosmological constant as positive pressure \cite{Kastor:2009wy,Cvetic:2010jb}. It should be emphasized that the first law is valid formally for all black hole horizons. Since the coupling constant $\alpha_2$ is dimensionful, the Smarr relation breaks down unless one wants to treat $\alpha_2$ as a thermodynamical variable.

Note that $\alpha_1=0$ is not a smooth limit, it alters the asymptotic behavior of electric potential $\phi'$, and hence it should be analysed separately.  In this case, the solution is given in (\ref{4spcases}), and the mass, electric and magnetic charges are
\be
M\,,\qquad Q_e=q\,,\qquad Q_m=p\,.
\ee
The other black hole thermodynamic quantities are
\bea
&&T=\fft{\epsilon}{4\pi r_+} - \fft{r_+}{4\pi}\big(\Lambda_0 + \fft{q^2}{4\alpha_2 p^2}\big)\,,\qquad
S=\pi r_+^2\,,\nn\\
&& \Phi_e= - \fft{q r_+^3}{12\alpha_2 p^2}\,,\qquad \Phi_m=\fft{q^2 r_+^3}{12 \alpha_2 p^2}\,.
\eea
It is easy to verify that the first law is satisfied.  Note that in this case we have $\Phi_e Q_e + \Phi_m Q_m=0$.

\section{Multiple photon spheres and a stable one}
\label{sec:ps}

In the curved spacetimes, it turns out that there can  exist the null geodesics circling around the black hole with a certain radius $\hat r$, giving rise to a photon sphere. For example, the radius of the photon sphere of the Schwarzschild black hole is $\fft32$ of the Schwarzschild radius \cite{Hod:2011aa,Hod:2012nk}. However, the photon sphere is unstable and hence has no observational consequence. Stable photon spheres are hard to come by and a wide search for such black holes in supergravities led no positive result \cite{Cvetic:2016bxi}. To our knowledge, there is hitherto no example of stable photon sphere in literature for asymptotically flat black holes in Einstein gravity with at least the null energy condition. In this section, we study the photon spheres in the dyonic black holes exhibited in the section \ref{subsec:dbh}. Especially, we show a stable photon sphere can indeed exists without violating the dominant energy condition.

In the spherically symmetric geometries, the radius $\hat r$ of a photon sphere are determined by
\be
\big(\fft{f}{r^2}\big)'\Big|_{r=\hat r}=0\,,\qquad\hbox{and}\qquad f(\hat r)>0\,.\label{pseq}
\ee
The second condition ensures that $r=\hat r$ is in the region where $r$ remains spacelike so that $r=\hat r$ describes a spatial 2-sphere. Whether the photon sphere is stable or not can be determined by the second derivative, namely \cite{Koga:2016jjq}
\bea
\hbox{stable}:&&\qquad \big(\fft{f}{r^2}\big)''\Big|_{r=\hat r}>0\,,\nn\\
\hbox{unstable}:&&\qquad \big(\fft{f}{r^2}\big)''\Big|_{r=\hat r}<0\,.
\eea
For example, for the Schwarzschild black hole with $f=1-2M/r$, it is easy to see that there is an unstable photon sphere located at $\hat r=3M$.

We now determine the photon spheres of the dyonic black holes presented in subsection \ref{subsec:dbh}. It is important to be reminded that all the black holes satisfy the null, weak and dominant energy conditions. First we consider the black hole given in (\ref{case1bh}).  The solution can have at most two horizons.  To be concrete, we choose $M=2$, in which case the horizons are located at
\be
r_1=3.0118\,,\qquad r_2=0.2846\,.
\ee
We find that there are two positive roots in $(f/r^2)'=0$, given by
\be
\tilde r_1=4.7366\,,\qquad \tilde r_2=0.3729\,.
\ee
We see that $f(\tilde r_1)>0$, but $f(\tilde r_2)<0$, it follows that there is only one photon sphere at $\hat r_1=\tilde r_1$, and it is unstable.

We now consider the solution (\ref{case2abh}) with $M=6.7$. In this case there are four real horizons, given by
(\ref{case2a4h}).  We find that there are four positive roots for $(f/r^2)'$:
\be
\tilde r_1=13.6955\,,\qquad
\tilde r_2=5.2044\,,\qquad
\tilde r_3=1.8311\,,\qquad
\tilde r_4=0.9428\,.
\ee
Compare with the horizon locations (\ref{case2a4h}), we conclude that there are only two photon spheres
\be
\hat r_1=\tilde r_1\,,\qquad \hat r_2=\tilde r_3\,.
\ee
Both photon spheres are unstable.  Note that the second photon sphere $\hat r_2$ is located at the inner spacetime region $r_3<r<r_2$, in which $r_3$ can be viewed as an event horizon and $r_2$ as the cosmic horizon.

We now study the photon spheres in the black hole (\ref{case2bbh}).  When the black hole has four horizons, the result is analogous to the above and there are two unstable photon spheres, one is in the outer world and the other is in the inner world.  Instead, we consider the two-horizon solution with $M=6.7$. The outer and inner horizons are given in (\ref{case2b2h}).  An unusual feature of this black hole is that the function $f$ is not monotonic outside the outer horizon, but has a wiggle between $(2.848,6.1421)$, such that $r=6.1421$ is a static equilibrium for a neutral massive particle.

We find that there are four positive roots for $(f/r^2)'$, given by
\be
\tilde r_1=12.4354\,,\qquad\tilde r_2=6.4430\,,\qquad
\tilde r_3=2.1939\,,\qquad \tilde r_4=0.8244\,.
\ee
In this case, we have 3 photon spheres
\be
\hat r_1=\tilde r_1\,\qquad \hat r_2=\tilde r_2\,,\qquad \hat r_3=\tilde r_3\,,
\ee
all of which are outside the outer horizon $r_1=1.5498$.  Furthermore, we find although the photon spheres at $\hat r_1$ and $\hat r_3$ are unstable, the photon sphere at $\hat r_2$ is {\it stable}. The result is suggestive of a connection between the existence of a stable static equilibrium for a neutral massive particle and a stable photon sphere.

To be complete, we now examine the photon spheres of the solution (\ref{case3bh}) with two extremal horizons. We find that there are again four positive roots for $(f/r^2)'$:
\be
\tilde r_1=6.716\,,\qquad
\tilde r_2=3\,,\qquad \tilde r_3=1.5075\,,\qquad \tilde r_4=1\,.
\ee
Again only $\hat r_1=\tilde r_1$ and $\hat r_2 = \tilde r_3$ are the radii of photon spheres. One is in the outer world while the other is in the inner world. Both of the photon spheres are unstable.

To conclude, depending on the choice of parameters, the asymptotically-flat black holes can have one, two or three photon spheres.  In the last case, one of the three is stable, providing the first such example in literature.  It should be emphasized that the black hole with stable photon sphere satisfies the dominant energy condition, but not the strong energy condition.

\section{On-shell action and the hidden electromagnetic duality}
\label{sec:hidden}

The free Maxwell theory has electromagnetic duality that interchanges its equation of motion and the Bianchi identity.  The symmetry remains when the source includes both electric and magnetic charges. However, in the standard formulation the symmetry breaks down in the on-shell action owing to the fact $F^2 = - (*F)^2$. (See \cite{Goldoni:2009zz,Deser:1976iy,Deser:1981fr,Deser:1996xu,Cremmer:1998px} for alternative formulations.)  The issue becomes recently urgent owing to the conjecture that the complexity of a quantum system is equal to the on-shell action of some AdS black hole in the Wheeler-DeWitt (WDW) patch \cite{ca1,ca2}. In other words, one would expect that the complexity in such a quantum system should have electromagnetic duality.

The issue was resolved for the Maxwell theory or a class of Einstein-Maxwel-dilaton theories by adding an appropriate boundary term at the price of introducing mixed boundary conditions in the variation principle \cite{goto}.  In \cite{Liu:2019smx}, the boundary term of a more general class of theories was proposed
\be
I=\fft{\gamma}{16\pi} \int_{\partial M} d\Sigma_\mu \widetilde  F^{\mu\nu} A_\nu\,,\qquad \widetilde F_{\mu\nu}
=-\fft{2\partial L}{\partial F_{\mu\nu}}\,.\label{genbound}
\ee
By examining several examples, it was conjectured that $\gamma=\ft12$ was the universal factor that could restore the electromagnetic duality of equations in the on-shell action.  An interesting question was asked in \cite{Liu:2019smx}: is there any significance of this boundary term when a theory does not have electromagnetic duality at all?

As we have seen, the quasi-topological term breaks the electromagnetic duality.  Furthermore the exact solution of general dyonic black holes can be constructed. These make our theory a perfect example to address the question raised in \cite{Liu:2019smx}.  It follows from \cite{Liu:2019smx} that the relevant boundary term is
\be
  I_{\mu Q} = \fft{\gamma}{4\pi} \, \int_{\partial M} d \Sigma_\mu  \big(  \alpha_1 F^{\mu\nu}  + 2 \alpha_2 F^2 F^{\mu\nu} - 4 \alpha_2 F^{\mu\rho}F^{\sigma}{}_{\rho} F_\sigma{}^\nu \big) A_\nu \,.
\ee
Making use of the equation of motion and the Stokes' theorem, the Maxwell boundary term on shell can be written as a bulk integration
\be
  I_{\mu Q}\Big|_{\text{on-shell}} =\fft{\gamma}{8\pi} \, \int_M dx^4 \sqrt g \big(  \alpha_1 F^2  + 2 \alpha_2 (F^2)^2 - 4 \alpha_2 F^4 \big) \,.
\ee
The dyonic black hole was presented in the previous section.  We can use the established technique \cite{Lehner:2016vdi} to evaluate the on-shell action growth rate, which includes four contributions, namely the bulk, Gibbons-Hawking surface term, the joint term and the $\gamma$ boundary term:
\be
\fft{d I}{dt} = \fft{d I_{\text{bulk}}}{dt} + \fft{d I_{\text{GH}}}{dt} + \fft{d I_{\text{joint}}}{dt} + \fft{d I_{\mu Q}}{dt} \,.
\ee
And we find that
\bea
\fft{d(I_{\text{bulk}} + I_{\text{Max}})}{dt} &=& \Big( Y(r) - \gamma Q_e \Phi_e - (1 - \gamma) Q_m \Phi_m \Big) \Big|_{r_-}^{r_+} \,, \cr
\fft{d( I_{\text{bd}} + I_{\text{joint}})}{dt} &=& - Y(r) \Big|_{r_-}^{r_+} \,.
\eea
The detail of the function $Y$ is not important since it cancels itself in the full action growth.  Nevertheless, we give the explicit expression
\be
Y = \frac{3 q^2 \, _2F_1\left(\frac{1}{4},1;\frac{5}{4};-\frac{4 p^2 \alpha_2}{r^4 \alpha_1}\right)}{4 \alpha_1 r}+\frac{2 \left(\epsilon r^2+2 \alpha_1 p^2\right) \left(4 \alpha_2 p^2+\alpha_1 r^4\right)+q^2 r^4}{4 \left(4 \alpha_2 p^2 r+\alpha_1 r^5\right)} \,.
\ee
Thus the total action growth rate can then be written as a simple form in terms of thermodynamical quantities as that of the Einstein-Maxwell case
\bea
\fft{d I}{d t} &=&\Big( - \gamma Q_e \Phi_e - (1 - \gamma) Q_m \Phi_m \Big) \Big|_{r_-}^{r_+}\nn\\
&=& -\ft12 \Big(\Phi_e Q_e+\Phi_m Q_m\Big) \Big|_{r_-}^{r_+}\,,\qquad\hbox{for}\qquad
\gamma=\ft12\,.\label{complexity}
\eea
When the black hole has only two horizons, $r_+$ and $r_-$ denote the outer and inner horizons.  When the black hole has four horizons, the definition of the WDW patch becomes less clear, and here we assume that it is region between $(r_2,r_1)$.  For the singular limit, $\alpha_1 = 0$, the solution  and thermodynamics were also presented in the previous section. Following the same procedure, we find that the late time action has the same form as (\ref{complexity}).

Thus we see that even though our quasi-topological electromagnetism breaks the electromagnetic duality at the level of equations, its on-shell action in the WDW patch has hidden electromagnetic duality when $\gamma=\fft12$, exactly the same as the dyonic RN-AdS black holes in Einstein-Maxwell theory.

\section{Quasi-topological electromagnetism as dark energy?}
\label{sec:darke}

In section \ref{sec:d=4}, we saw that the quasi-topological term associated with $\alpha_2$ gave rise to a perfect-fluid energy-momentum tensor that resembled the dark energy.  In this section, we demonstrate the reason behind and also study its application in cosmology.  We first consider the simplest model
\be
{\cal L}=\sqrt{-g} \Big( R -\alpha_2 \big((F^2)^2 - 2F^{(4)}\big)\Big)\,.
\ee
It is easy to verify that the quasi-topological term is effectively a positive cosmological constant on shell.
To see this we note that
\be
((F^2)^2 - 2F^{(4)}) {*\oneone}\, \sim\, {*(F\wedge F)}\wedge F\wedge F\,.
\ee
In $D=4$, we must have ${*(F\wedge F)}=2\psi(x)$ for certain function $\psi$.  The equation of motion associated with the variation of $A$ is then given by
\be
d(\psi\, F)=0\,.
\ee
Thus $\psi$ must be a constant.  Assuming that $\psi=\lambda$, the quasi-topological term is then effectively the cosmological constant with
\be
\Lambda_{\rm eff}=4 \alpha_2 \lambda^2\,.
\ee
Thus we can construct the cosmological de Sitter spacetime
\be
ds^2 = -dt^2 + a(t)^2 (dx^2 + dy^2 + dz^2)\,,\qquad a=e^{2\sqrt{\alpha_2/3}\, \lambda\, t}\,.
\ee
We see that in this model, the cosmological constant is a composite quantity, built from the fundamental $U(1)$ field $A$.  In this particular example, the solution for $A$ can be
\bea
&&A = \big(x \psi_1 (t) + y \psi_2(t) + z \psi_3 (t)\big) dt +
p_1 y dz + p_2 z dx + p_3 x dy\,,\nn\\
&&p_1 \psi_1 + p_2 \psi_2 + p_3 \psi_3 = \lambda a^3\,.
\eea
Even in this simple model, there is an important difference between quasi-topological electromagnetism and a fixed cosmological constant.  In the former the effective cosmological constant is an integration constant of the theory rather than a fixed value as in the latter case.

The true advantage of having a composite model for the dark energy is that it allows to have interesting couplings between the dark energy and other matter fields.  We now present a concrete model with a nontrivial coupling between the dark energy and a scalar field $\phi$.  The Lagrangian is given by
\be
{\cal L}=\sqrt{-g} \Big( R -\ft12 (\partial\phi)^2 -V(\phi) -\alpha_2 \big((F^2)^2 - 2F^{(4)}\big)\Big)
- \ft12 U(\phi) \varepsilon^{\mu\nu\rho\sigma}
F_{\mu\nu} F_{\rho\sigma}\,.
\ee
Here $\varepsilon^{\mu\nu\rho\sigma}$ is totally antisymmetric density with $\varepsilon^{0123}=1$. Defining
tensor $\epsilon^{\mu\nu\rho\sigma}=\varepsilon^{\mu\nu\rho\sigma}/\sqrt{-g}$, the covariant equations of motion are
\bea
&&\Box \phi=\fft{\partial V}{\partial\phi} + \fft12 \fft{\partial U}{\partial \phi} \epsilon^{\mu\nu\rho\sigma}
F_{\mu\nu} F_{\rho\sigma}\,,\qquad \nabla_\mu \widetilde F^{\mu\nu} =2\fft{\partial U}{\partial \phi} \epsilon^{\nu\mu\rho\sigma} \partial_\mu \phi F_{\rho\sigma}\,,\nn\\
&&R_{\mu\nu} - \ft12 g_{\mu\nu} = \ft12 (\partial_\mu \phi\partial_\nu \phi - \ft12 (\partial\phi)^2 g_{\mu\nu}) - \ft12 V(\phi) g_{\mu\nu} + \alpha_2 T^{(2)}_{\mu\nu}\,.
\eea
We find that the theory admits the Friedmann-Lema\^ itre-Robertson-Walker (FLRW) cosmological solution, with the ansatz
\bea
ds^2 &=& -dt^2 + a(t)^2 (dx^2 + dy^2 + dz^2)\,,\nn\\
A &=& \ft13 (x + y + z) a(t)^3 \psi(t) dt + p (y dz +  z dx + x dy)\,,\qquad \phi=\phi(t)\,.
\eea
Note that we have ${* (F\wedge F)}=2p\psi(t)$. The equations of motion for the scalar and $A$ fields are
\be
-\ddot \phi - \fft{3\dot a}{a} \dot \phi =- 4 p \psi \fft{\partial U}{\partial \phi} + \fft{\partial V}{\partial \phi}\,,\qquad \alpha_2 p \dot \psi = \ft1{4} \dot U\,.\label{mattereom}
\ee
The Einstein equations give the standard cosmological equations
\be
\fft{\ddot a}{a} = - \fft 1 {12} ( \rho + 3 p ) \,, \qquad \fft{3\dot a^2}{a^2}=2\rho \,,
\ee
with
\be
\rho= \ft12 {\dot \phi^2} + V + 8\alpha_2 p^2\psi^2 \,,\qquad
p=\ft12 \dot \phi^2 - V -8 \alpha_2 p^2 \psi^2\,.
\ee
Since our theory is constructed from the Lagrangian formalism using fundamental fields, the law of conservation of energy-momentum tensor
\be
\dot \rho + 3 \fft{\dot a}{a} (\rho + p) = 0\,,
\ee
is automatically satisfied.  Note that the second equation of (\ref{mattereom}) can be solved exactly, given by
\be
\psi = q (1 + \fft{U}{4\alpha_2 pq})\,.
\ee
It follows that resulting equations of motion are the same as the FLRW cosmology driven by a scalar $\phi$ with effective scalar potential
\be
V_{\rm eff}(\phi) = V(\phi)+ 8 \alpha_2 p^2 q^2 (1 + \fft{U(\phi)}{4\alpha_2 pq})^2\,.
\ee
As a concrete and simple example, we take $V=0$ and $U=\beta\phi^2\ge 0$.  In this case, if the constant flux $pq>0$, then $\phi=0$ is a stable de Sitter vacuum and the effective cosmological constant $\Lambda_{\rm eff} = 4\alpha_2 p^2 q^2$ should describe later time dark energy.  On the other hand, if $p q<0$, $\phi=0$ with $\Lambda_{\rm eff}$ is the unstable de Sitter vacuum and hence the model can be used to describe the early inflation.

By the nature of the construction, our theory can be reduced to an effective scalar model in cosmology. The $U(1)$ vector can of course have more versatile couplings with other matter. In the above construction, the $U(1)$ vector is not the usual Maxwell field of photons.  It is tantalizing to speculate the possibility that the quasi-topological term is indeed the extension of the Maxwell field. A single photon will give no contribution to the energy density $(\vec E\cdot \vec B)^2$, but the random radiation in the cosmic microwave background can contribute and give rise to
\be
\rho=-p \sim  \fft{\rho_{\rm photon}^2}{M}\,,
\ee
where $M$ is some fundamental scale required by dimensional consistency. This quantity is clearly too small to account for the current dark energy.  On the other hand, in the early universe where the radiation energy is large, it can be nontrivial and provide a driving force for inflation.  However, by itself, it is unlikely to create enough e-foldings since the photon density reduces quickly with the inflation and hence $\rho$ has the wrong falloff with respect to the cosmological scaling factor $a$.  If the quasi-topological electromagnetism is the higher-order extension of the Maxwell theory, the radiation is unlikely the composite for the dark energy. An alternative possibility is magnetic monopoles, which, together the electric charges, can contribute in part $\rho=-p$.  As we have seen in the previous sections, the asymptotic spacetime is flat if there is no additional cosmological constant.  However, in this case, there can be an inner world sandwiched between the cosmic horizon and black hole horizons. Its implication in cosmology requires further exploration.

\section{Conclusions}
\label{sec:conclusion}

We introduced a new concept of quasi-topological electromagnetism which is defined as the squared norm of the topological wedge products of the Maxwell field strength in general $(k\ge 2)$'th order. (The $k=1$ leads to the standard kinetic term of the Maxwell theory.) We focus on the study in four spacetime dimensions, in which case, we have $k=2$ only.  One salient property of the quasi-topological term in $D=4$ is that it contributes an energy-momentum tensor of isotropic perfect fluid, namely $T^{ab}={\rm diag} \{\rho,-\rho,-\rho,-\rho\}$, with $\rho\propto (\vec E\cdot \vec B)^2$. It can thus be used as a model for composite dark energy where a vector $U(1)$ field is its fundamental ingredient.  We also found that it could provide an {\it on-shell} spontaneous symmetry breaking mechanism where a massless scalar could acquire a mass with the sign of the mass square depending on the sign of the $F\wedge F$ flux.

We first considered Einstein-Maxwell theory extended with this quasi-topological term and studied its application in static black hole physics. The matter sector satisfies the dominant energy condition, and hence also the null and weak energy conditions, but it can violate the strong energy condition. By the nature of the construction, the RN black holes with either electric or magnetic charges remain exact solutions, but the dyonic black holes are modified and they exhibit many unusual properties. We found that the dyonic black holes were all Schwarzschild-like, (i.e.~$g_{tt} g_{rr}=-1$), with the blackening factor $f$ being expressed in terms of a hypergeometric function.  The solutions carry three independent parameters, the mass, the electric and magnetic charges.

As in the case of RN black holes, the black hole singularity at $r=0$ is time-like and hence there are an even number of horizons. (An extremal horizon is counted as two here.) For some fixed choices of electric and magnetic charges, adjusting the mass parameter can give at most two horizons.  These black holes are analogous to the RN solutions, satisfying both the dominant and strong energy conditions. For some other choices of electric and magnetic charges, adjusting the mass can yield two or even four black hole horizons $r_1>r_2>r_3>r_4>0$.  These black holes satisfy the dominant energy condition, but not the strong energy condition. It follows that the region $r\in (r_3,r_2)$ is like regular spacetime sandwiched between a black hole horizon and a cosmic horizon.  In other words, there is an ``inner new world'' inside the dyonic black hole. We also study the photon spheres created by the dyonic black hole and we find that there is a photon sphere in both inner and outer worlds, but both are unstable.

When a black hole that could have four horizons, but it will reduce to two horizons if we lower the mass appropriately. Then new phenomena can arise. For suitable parameters, we found that the blackening function $f$ outside the outer horizon could be non-monotonic with a local maximum followed by a local minimum before increasing monotonically to asymptotic infinity.  This implies that gravity is repulsive in the region between the maximum and the minimum, giving rise to a stable static equilibrium at the local minimum. In this case, we found that there existed three photon spheres, one located between the outer horizon and the local maximum and two located beyond the local minimum.  The most inner and outer photon spheres are unstable, but the middle one is {\it stable}, providing the first example of stable photon sphere in the asymptotically flat spacetime geometry that satisfies at least the null energy condition.

The quasi-topological term breaks the electromagnetic duality and the electric and magnetic charges $(Q_e,Q_m)$ enter the dyonic black holes asymmetrically.  However, we find that by adding the boundary term (\ref{genbound}), the on-shell action of the dyonic black holes has hidden electromagnetic duality at precisely the $\gamma=\ft12$ value advocated in \cite{Liu:2019smx}. This is indicative that this hidden symmetry could be universal in higher-order extended Maxwell theories.

We finally studied applications of the quasi-topological electromagnetism in cosmology. We focused on the case where the quasi-topological term is not an extension of the Maxwell theory.  When the term is on its own, we showed that it gave rise an effective cosmological constant. We then considered the coupling of the fundamental $U(1)$ ingredient with some massless scalar field and obtained an effective potential with non-vanishing mass.  The sign of the mass squared depends on the sign of the flux $F\wedge F$.  By the nature of construction, this provides an FLRW (homogeneous and isotropic) cosmological model that may have applications in early cosmology or explaining later time dark energy. The key advantage of the quasi-topological term as being the dark energy is that its fundamental ingredient $A_\mu$ has a variety of ways to couple to other forms of matter, including the dark matter.

Our introduction of the quasi-topological electromagnetism and our initial exploration turn out to reveal many intriguing properties in both black hole physics and cosmology.  Each of these properties deserves further investigation. We have focused on the quasi-topological polynomials for the $U(1)$ field. It is straightforward to generalize to involve the Yang-Mills field strength as well, namely $|{\rm tr} (F\wedge F)|^2$, (or even a generic function of this quantity.) It is of great interest to investigate its implications.

\section*{Acknowledgement}

We are grateful to Qing-Guo Huang, Pu-Jian Mao, Zhao-Long Wang and Jun-Bao Wu for useful discussions.
H.-S.L.~is supported in part by NSFC (National Natural Science Foundation of China) Grants No.~11475148 and No.~11675144. Y.-Z.L., Z.-F.M.~and H.L.~are supported in part by NSFC Grants No.~11875200 and No.~11475024.

\appendix

\section{Dyonic black holes in higher even dimensions}
\label{sec:app}

In this appendix, we consider the Lagrangian (\ref{d4lag}) in general $D=2n+2$ dimensions and construct
dyonic black holes analogous to the ones in the Einstein-Born-Infeld theory \cite{Li:2016nll}. The ansatz for the dyonic black hole in even dimensions is
\bea
ds^2&=&-f(r)dt^2+\frac{dr^2}{f(r)}+r^2(d\Omega_{1,\epsilon}^2+d\Omega_{2,\epsilon}^2+\cdots d\Omega_{n,\epsilon}^2)\,,\nn\\
F &=& -\phi'(r)dt\wedge dr+p (\Omega_1 + \Omega_2 + \cdots + \Omega_n)\,.
\eea
where
\be
\phi'(r)=\frac{q r^{6-D}}{\alpha_1 r^4+ 2(D-2) \alpha_2 p^2}
\ee
It can be solved that
\be
\phi(r)=\frac{q}{\alpha_1 r^{D-3}}\,_2F_1\Big(1,\frac{1}{4}(D-3);\frac{D+1}{4};
-\frac{2(D-2) \alpha_2 p^2}{\alpha_1 r^4} \Big)\,.
\ee
the equation of motion with respect to $f$ is given by:
\be
f'+\fft{(D-3)f-\epsilon}{r}+ \fft{\alpha_1 p^2}{r^3}+\fft{(D-4)\alpha_2 p^4}{r^7}
-\fft{2qa'}{(D-2)r^{D-3}}+ \fft{2\Lambda_0}{D-2}r=0
\ee
The function $f$ can be solved as
\bea
f=&&-\fft{2\Lambda_0 r^2}{(D-1)(D-2)}+\fft{\epsilon}{D-3}-\fft{\mu}{r^{D-3}}-\fft{\alpha_1 p^2}{(D-5)r^2}-\fft{(D-4)\alpha_2 p^4}{(D-9)r^6}\cr
&&+\fft{2q^2}{(D-2)(D-3)\alpha_1}\fft{1}{r^{2(D-3)}}\,
_2F_{1}(1;\ft{1}{4}(D-3);\ft{1}{4}(D+1);-\fft{2(D-2)\alpha_2 p^2}{\alpha_1 r^4}).
\eea

The mass and electric and magnetic charges of the solution are
\bea
M&=&\frac{n\omega_2^{n}}{8\pi}\mu\,,\nn\\
Q_e=\frac{\omega_2^n}{16\pi}\int r^{2n}h^{tr}|_{r\ra \infty}=\frac{ q}{16\pi}\omega^n_2\,, &&\qquad Q_m=\frac{\omega_2^n}{16\pi}\int F_{\theta \phi}|_{r \ra \infty}=\frac{p}{16\pi}\omega_2^n\,.
\eea
For sufficiently large $M$, the solutions describe black holes with the event horizon being the largest root $r_+$ of $f$. The temperature and the entropy of these black holes are
\be
T=\frac{f'(r_+)}{4\pi}\,,\qquad S=\frac{r_+^{2n}}{4}\omega_2^{n}\,.
\ee
The electric and magnetic potentials are
\bea
\Phi_e&&=\frac{q}{4\alpha_1 r_+^{D-3}}\,_2F_1\Big(1,\frac{1}{4}(D-3),\frac{D+1}{4},-\frac{2(D-2)\alpha_2 p^2 }{\alpha_1 r_+^4} \Big)\,,\nn\\
\Phi_m&&=\frac{2(D-2)\alpha_1 p r_+^{D-5}}{(D-5)}-\frac{4(D-4)(D-2) \alpha_2 p^3 r_+^{D-9}}{D-9}+\frac{q^2}{(r_+^4 \alpha_1+2(D-2)\alpha_2 p^2)r_+^{D-7}}\nn\\
&&-\frac{q^2}{r_+^{D-3}}\, _2F_1\Big(1;\frac{D-3}{4};\frac{D+1}{4};-\frac{2(D-2)\alpha_2p^2}{\alpha_1r_+^4 }\Big)\,.
\eea
These thermodynamical quantities satisfy the first law
\be
dM=TdS+\Phi_e d Q_e +\Phi_m d Q_m + \frac{r_+^{D-1}\omega_2^n}{(D-1)}\, d \big(-\frac{\Lambda_0}{8\pi}\big).
\ee


\begin{thebibliography}{99}

\bibitem{Born:1934gh}
  M.~Born and L.~Infeld,
  ``Foundations of the new field theory,''
  Proc.\ Roy.\ Soc.\ Lond.\ A {\bf 144}, no. 852, 425 (1934).
  doi:10.1098/rspa.1934.0059

\bibitem{Oliva:2010eb}
  J.~Oliva and S.~Ray,
  ``A new cubic theory of gravity in five dimensions: Black hole, Birkhoff's theorem and C-function,''
  Class.\ Quant.\ Grav.\  {\bf 27}, 225002 (2010)
  doi:10. 1088/0264-9381/27/22/225002
  [arXiv:1003.4773 [gr-qc]].

\bibitem{Myers:2010ru}
  R.C.~Myers and B.~Robinson,
``Black holes in quasi-topological gravity,''
  JHEP {\bf 1008}, 067 (2010)
  doi:10.1007/JHEP08(2010)067
  [arXiv:1003.5357 [gr-qc]].

\bibitem{Dehghani:2011vu}
  M.H.~Dehghani, A.~Bazrafshan, R.B.~Mann, M.R.~Mehdizadeh, M.~Ghanaatian and M.H.~Vahidinia,
``Black holes in quartic quasitopological gravity,''
  Phys.\ Rev.\ D {\bf 85}, 104009 (2012)
  doi:10.1103/PhysRevD.85.104009
  [arXiv:1109.4708 [hep-th]].

\bibitem{Li:2017ncu}
  Y.Z.~Li, H.S.~Liu and H.~L\"u,
  ``Quasi-topological Ricci polynomial gravities,''
  JHEP {\bf 1802}, 166 (2018)
  doi:10.1007/JHEP02(2018)166
  [arXiv:1708.07198 [hep-th]].

\bibitem{Wang:2016lxa}
  B.~Wang, E.~Abdalla, F.~Atrio-Barandela and D.~Pavon,
  ``Dark matter and dark energy interactions: theoretical challenges, cosmological implications and observational signatures,''
  Rept.\ Prog.\ Phys.\  {\bf 79}, no. 9, 096901 (2016)
  doi:10.1088/0034-4885/79/9/096901
  [arXiv:1603.08299 [astro-ph.CO]].

\bibitem{Horndeski:1976gi}
  G.W.~Horndeski,
  ``Conservation of charge and the Einstein-Maxwell field equations,''
  J.\ Math.\ Phys.\  {\bf 17}, 1980 (1976).
  doi:10.1063/1.522837

\bibitem{Goldoni:2009zz}
  O.~Goldoni and M.F.A.~da Silva,
  ``Energy conditions for electromagnetic field in presence of cosmological constant,''
  PoS ISFTG {\bf }, 072 (2009).
  doi:10.22323/1.081.0072

\bibitem{Gao:2017vqv}
  C.~Gao, Y.~Lu, S.~Yu and Y.G.~Shen,
  ``Black hole and cosmos with multiple horizons and multiple singularities in vector-tensor theories,''
  Phys.\ Rev.\ D {\bf 97}, no. 10, 104013 (2018)
  doi:10.1103/PhysRevD.97.104013
  [arXiv:1711.00996 [gr-qc]].

\bibitem{Horne:1992bi}
  J.H.~Horne and G.T.~Horowitz,
  ``Black holes coupled to a massive dilaton,''
  Nucl.\ Phys.\ B {\bf 399}, 169 (1993)
  doi:10.1016/0550-3213(93)90621-U
  [hep-th/9210012].

\bibitem{Lee:1994sk}
  K.M.~Lee and E.J.~Weinberg,
  ``Nontopological magnetic monopoles and new magnetically charged black holes,''
  Phys.\ Rev.\ Lett.\  {\bf 73}, 1203 (1994)
  doi:10.1103/PhysRevLett. 73.1203
  [hep-th/9406021].

\bibitem{Gibbons:1994ff}
  G.W.~Gibbons and R.E.~Kallosh,
  ``Topology, entropy and Witten index of dilaton black holes,''
  Phys.\ Rev.\ D {\bf 51}, 2839 (1995)
  doi:10.1103/PhysRevD.51.2839
  [hep-th/9407118].

\bibitem{Lee:1991qs}
  K.M.~Lee, V.P.~Nair and E.J.~Weinberg,
  ``A Classical instability of Reissner-Nordstrom solutions and the fate of magnetically charged black holes,''
  Phys.\ Rev.\ Lett.\  {\bf 68}, 1100 (1992)
  doi:10.1103/PhysRevLett.68.1100
  [hep-th/9111045].

\bibitem{Zhao:2018wkl}
  Q.Q.~Zhao, Y.Z.~Li and H.~L\"u,
  ``Static equilibria of charged particles around charged black holes: chaos bound and its violations,''
  Phys.\ Rev.\ D {\bf 98}, no. 12, 124001 (2018)
  doi:10.1103/PhysRevD.98.124001
  [arXiv:1809.04616 [gr-qc]].

\bibitem{Lu:2019icm}
  H.~L\"u, Z.L.~Wang and Q.Q.~Zhao,
  ``Black holes that repel,''
  Phys.\ Rev.\ D {\bf 99}, no. 10, 101502 (2019)
  doi:10.1103/PhysRevD.99.101502
  [arXiv:1901.02894 [hep-th]].

\bibitem{Kastor:2009wy}
  D.~Kastor, S.~Ray and J.~Traschen,
``Enthalpy and the mechanics of AdS black holes,''
  Class.\ Quant.\ Grav.\  {\bf 26}, 195011 (2009)
  doi:10.1088/0264-9381/26/19/195011
  [arXiv:0904.2765 [hep-th]].

\bibitem{Cvetic:2010jb}
  M.~Cveti\v c, G.W.~Gibbons, D.~Kubiznak and C.N.~Pope,
``Black hole enthalpy and an entropy inequality for the thermodynamic volume,''
  Phys.\ Rev.\ D {\bf 84}, 024037 (2011)
  doi:10.1103/PhysRevD.84.024037
  [arXiv:1012.2888 [hep-th]].

\bibitem{Hod:2011aa}
  S.~Hod,
  ``Hairy black holes and null circular geodesics,''
  Phys.\ Rev.\ D {\bf 84}, 124030 (2011)
  doi:10.1103/PhysRevD.84.124030
  [arXiv:1112.3286 [gr-qc]].

\bibitem{Hod:2012nk}
  S.~Hod,
  ``The fastest way to circle a black hole,''
  Phys.\ Rev.\ D {\bf 84}, 104024 (2011)
  doi:10.1103/PhysRevD.84.104024
  [arXiv:1201.0068 [gr-qc]].

\bibitem{Cvetic:2016bxi}
  M.~Cveti\v c, G.W.~Gibbons and C.N.~Pope,
 ``Photon spheres and sonic horizons in black holes from supergravity and other theories,''
  Phys.\ Rev.\ D {\bf 94}, no. 10, 106005 (2016)
  doi:10.1103/PhysRevD.94.106005
  [arXiv:1608.02202 [gr-qc]].

\bibitem{Koga:2016jjq}
  Y.~Koga and T.~Harada,
  ``Correspondence between sonic points of ideal photon gas accretion and photon spheres,''
  Phys.\ Rev.\ D {\bf 94}, no. 4, 044053 (2016)
  doi: 10.1103/PhysRevD.94.044053
  [arXiv:1601.07290 [gr-qc]].

\bibitem{Deser:1976iy}
  S.~Deser and C.~Teitelboim,
``Duality transformations of abelian and nonabelian gauge fields,''
  Phys.\ Rev.\ D {\bf 13}, 1592 (1976).
  doi:10.1103/PhysRevD.13.1592

\bibitem{Deser:1981fr}
  S.~Deser,
  ``Off-shell electromagnetic duality invariance,''
  J.\ Phys.\ A {\bf 15}, 1053 (1982).
  doi:10.1088/0305-4470/15/3/039

\bibitem{Deser:1996xu}
  S.~Deser, M.~Henneaux and C.~Teitelboim,
  ``Electric-magnetic black hole duality,''
  Phys.\ Rev.\ D {\bf 55}, 826 (1997)
  doi:10.1103/PhysRevD.55.826
  [hep-th/9607182].

\bibitem{Cremmer:1998px}
  E.~Cremmer, B.~Julia, H.~L\"u and C.N.~Pope,
  ``Dualization of dualities. 2. Twisted self-duality of doubled fields, and superdualities,''
  Nucl.\ Phys.\ B {\bf 535}, 242 (1998)
  doi:10.1016/S0550-3213(98)00552-5
  [hep-th/9806106].

\bibitem{Li:2016nll}
  S.~Li, H.~L\"u and H.~Wei,
  ``Dyonic (A)dS black holes in Einstein-Born-Infeld theory in diverse dimensions,''
  JHEP {\bf 1607}, 004 (2016)
  doi:10.1007/JHEP07(2016)004
  [arXiv:1606.02733 [hep-th]].

\bibitem{ca1}
  A.R.~Brown, D.A.~Roberts, L.~Susskind, B.~Swingle and Y.~Zhao,
 ``Holographic complexity equals bulk action?''
  Phys.\ Rev.\ Lett.\  {\bf 116}, no. 19, 191301 (2016), arXiv:1509. 07876 [hep-th].

\bibitem{ca2}
  A.R.~Brown, D.A.~Roberts, L.~Susskind, B.~Swingle and Y.~Zhao,
  ``Complexity, action, and black holes,''
  Phys.\ Rev.\ D {\bf 93}, no. 8, 086006 (2016), arXiv:1512.04993 [hep-th].

\bibitem{goto}
  K.~Goto, H.~Marrochio, R.C.~Myers, L.~Queimada and B.~Yoshida,
  ``Holographic complexity equals which action?''
  JHEP {\bf 1902}, 160 (2019),
  arXiv:1901.00014 [hep-th].

\bibitem{Liu:2019smx}
  H.S.~Liu and H.~L\"u,
  ``Action growth of dyonic black holes and electromagnetic duality,''
  arXiv:1905.06409 [hep-th].

\bibitem{Lehner:2016vdi}
  L.~Lehner, R.C.~Myers, E.~Poisson and R.D.~Sorkin,
 ``Gravitational action with null boundaries,''
Phys.\ Rev.\ D {\bf 94}, no. 8, 084046 (2016), arXiv:1609.00207 [hep-th].


\end{thebibliography}
\end{document}